\documentclass[12pt,manuscript]{emulateapj}

\begin{document}

\title{CANGAROO-III search for TeV Gamma-rays from two clusters of galaxies}

\author{R. Kiuchi\altaffilmark{1}, 
M. Mori\altaffilmark{1}, 
G. V. Bicknell\altaffilmark{2}, 
R. W. Clay\altaffilmark{3}, 
P. G. Edwards\altaffilmark{4}, 
R. Enomoto\altaffilmark{1},   \\
S. Gunji\altaffilmark{5},
S. Hara\altaffilmark{6}, 
T. Hara\altaffilmark{7},
T. Hattori\altaffilmark{8},  
S. Hayashi\altaffilmark{9},  
Y. Higashi\altaffilmark{10},  
Y. Hirai\altaffilmark{11},   \\
K. Inoue\altaffilmark{5},  
C. Itoh\altaffilmark{6},  
S. Kabuki\altaffilmark{10},  
F. Kajino\altaffilmark{9},  
H. Katagiri\altaffilmark{12}, 
A. Kawachi\altaffilmark{8},   
T. Kifune\altaffilmark{1},  \\
H. Kubo\altaffilmark{10}, 
J. Kushida\altaffilmark{8}, 
Y. Matsubara\altaffilmark{13},  
T. Mizukami\altaffilmark{10},  
Y. Mizumoto\altaffilmark{14},  
R. Mizuniwa\altaffilmark{8},   \\
H. Muraishi\altaffilmark{15},  
Y. Muraki\altaffilmark{13},  
T. Naito\altaffilmark{7}, 
T. Nakamori\altaffilmark{16},  
S. Nakano\altaffilmark{10},  
D. Nishida\altaffilmark{10},  \\
K. Nishijima\altaffilmark{8},  
M. Ohishi\altaffilmark{1},    
Y. Sakamoto\altaffilmark{8},  
A. Seki\altaffilmark{8},       
V. Stamatescu\altaffilmark{3},  
T. Suzuki\altaffilmark{11},    \\
D. L. Swaby\altaffilmark{3},  
T. Tanimori\altaffilmark{10},  
G. Thornton\altaffilmark{3},  
F. Tokanai\altaffilmark{5},  
K. Tsuchiya\altaffilmark{17},  
S. Watanabe\altaffilmark{10},  \\
Y. Yamada\altaffilmark{9},  
E. Yamazaki\altaffilmark{8},  
S. Yanagita\altaffilmark{11},  
T. Yoshida\altaffilmark{11},  
T. Yoshikoshi\altaffilmark{1}, and  
Y. Yukawa\altaffilmark{1} 
}

\altaffiltext{1}{Institute for Cosmic Ray Research, University of Tokyo, Kashiwa, Chiba 277-8582, Japan}
\altaffiltext{2}{Research School of Astronomy and Astrophysics, Australian National University, ACT 2611, Australia}
\altaffiltext{3}{School of Chemistry and Physics, University of Adelaide, SA 5005, Australia}
\altaffiltext{4}{CSIRO Australia Telescope National Facility, Narrabri, NSW
2390, Australia}
\altaffiltext{5}{Department of Physics, Yamagata University, Yamagata, Yamagata 990-8560, Japan}
\altaffiltext{6}{Ibaraki Prefectural University of Health Sciences, Ami, Ibaraki 300-0394, Japan} 
\altaffiltext{7}{Faculty of Management Information, Yamanashi Gakuin University, Kofu, Yamanashi 400-8575, Japan}
\altaffiltext{8}{Department of Physics, Tokai University, Hiratsuka, Kanagawa 259-1292, Japan} 
\altaffiltext{9}{Department of Physics, Konan University, Kobe, Hyogo 658-8501, Japan}
\altaffiltext{10}{Department of Physics, Kyoto University, Sakyo-ku, Kyoto 606-8502, Japan} 
\altaffiltext{11}{Faculty of Science, Ibaraki University, Mito, Ibaraki 310-8512, Japan} 
\altaffiltext{12}{Department of Physical Science, Hiroshima University, Higashi-Hiroshima, Hiroshima 739-8526, Japan} 
\altaffiltext{13}{Solar-Terrestrial Environment Laboratory,  Nagoya University, Nagoya, Aichi 464-8602, Japan} 
\altaffiltext{14}{National Astronomical Observatory of Japan, Mitaka, Tokyo 181-8588, Japan} 
\altaffiltext{15}{School of Allied Health Sciences, Kitasato University, Sagamihara, Kanagawa 228-8555, Japan} 
\altaffiltext{16}{Department of Basic Physics, Tokyo Institute of Technology, Meguro, Tokyo 152-8551, Japan}
\altaffiltext{17}{National Research Institute of Police Science, Kashiwa, Chiba 277-0882, Japan}

\begin{abstract}
Because accretion and merger shocks in clusters of galaxies may accelerate particles to high energies, 
clusters are candidate sites for the origin of ultra-high-energy (UHE) cosmic-rays.
A prediction was presented for gamma-ray emission from a cluster of galaxies at a 
detectable level with 
the current generation of 
imaging atmospheric Cherenkov telescopes. 
The gamma-ray emission was produced 
via inverse Compton upscattering of cosmic microwave background (CMB) photons 
by electron-positron pairs generated by collisions of UHE cosmic rays in the cluster.  
We observed two clusters of galaxies, Abell 3667 and Abell 4038, 
searching for very-high-energy gamma-ray emission with 
the CANGAROO-III atmospheric Cherenkov telescope system in 2006.
The analysis showed no significant excess around these clusters, 
yielding upper limits on the gamma-ray emission. 
From a comparison of the upper limit for the north-west radio relic region of Abell 3667
with a model prediction, 
we derive a lower limit for the magnetic field of the region
of ${\sim}0.1{\mu}$G. 
This shows the potential of gamma-ray observations in studies
of the cluster environment.
We also discuss the flux upper limit from cluster center regions using
a model of gamma-ray emission from neutral pions produced in hadronic collisions of 
cosmic-ray protons with the intra-cluster medium (ICM).
The derived upper limit of the cosmic-ray energy density within this framework 
is an order of magnitude higher than that of our Galaxy.

\end{abstract}


\section{Introduction}
\label{sec_intro}
Clusters of galaxies are the largest systems in the Universe that are gravitationally bound, and 
they are potential sources of ultra-high-energy (UHE) cosmic rays, since their sizes and 
moderate magnetic fields allow a high maximum energy (${\sim}10^{20}$\,eV) 
in acceleration \citep{Ostrowski_Hillas_diagram}. 
Although cluster accretion and merger shocks could produce such high-energy particles, 
accretion shocks may be more effective than merger shocks in particle acceleration,
due to their high Mach numbers \citep{Miniati,Ryu2003}. 
Cosmic-ray electrons accelerated directly by these shocks may produce gamma-ray emission
via inverse Compton (IC) scattering of the cosmic microwave background (CMB)
\citep{Totani_Kitayama,Miniati_2003,Gabici2004}.
On the other hand, accelerated cosmic ray protons can interact hadronically with 
the intra-cluster medium (ICM), and gamma-rays may be produced via ${\pi}^{0}$-decay 
\citep{Volk_1996,Berezinsky,Pfrommer_Ensslin} 
as well as IC emission by secondary electron/positron pairs from
${\pi}^{\pm}$-decay \citep{Blasi_secondary_1999}. 

Observations of clusters of galaxies at various wavelengths
(e.g., radio, EUV, X-ray) suggest the existence of 
non-thermal particles in these gigantic objects \citep{Fusco-Femiano_2001,Xray_summary,Bowyer_EUV,Radio_relic_summary}.
However, at gamma-ray energies, no observational evidence has been reported 
from clusters of galaxies \citep{Reimer_EGRET},
though there is suggestive evidence \citep{Kawasaki_Totani,Scharf}.
Observations in the TeV energy band with Imaging Atmospheric Cherenkov
Telescope (hereafter IACT) experiments have
yielded the only upper limits to date \citep{Hattori_ICRC,Fegan,Perkins,HESS_COMA_ICRC}.  

Recently, \citet{Inoue_cluster} considered protons, accelerated up to
$10^{18}$--$10^{19}$ eV by 
accretion shocks around a massive cluster, 
interacting with the CMB photons,
with secondary electron-positron pairs produced in the $p$-$\gamma$ process, 
boosting those photons into the TeV energy range by IC scattering.
Although their prediction depends on many physical parameters, 
the predicted gamma-ray flux could be 
at a detectable level for current IACT experiments for massive, nearby clusters. 
Thus, observations of clusters of galaxies with IACTs 
probe high-energy processes and the environment in these large-scale systems,
and if gamma-ray signals are detected, they may also provide
clues to help solve the mystery of UHE cosmic-ray production.
If there is no detected signal, we can place
limits on the physical parameters of clusters, such as the strength of 
the magnetic field.

In this paper we report on a search for TeV gamma-ray emission from two clusters of galaxies,
Abell 3667 and Abell 4038, with CANGAROO-III, 
an array of imaging atmospheric Cherenkov telescopes.
We selected these targets according to their high masses and relative
closeness 
from the southern Abell catalog \citep{Abell_catalog}.  

Abell 3667, also known as SC 2009-57, is classified as type~L in the 
Rood-Sastry system \citep{Rood_Sastry} due to the linear arrangement of the galaxies,
including two of the brightest D galaxies.
The cluster has a redshift of $z=0.055$ \citep{Sodre_A3667}, 
and is centered at $[\alpha(2000)=20^h 12^m 27^{s}.4, \delta(2000)=-56^{\circ}49'36"]$,
which is the location of the brightest D galaxy \citep{Knopp_xray_A3667}. 
The cluster is one of the brightest X-ray sources in the southern sky \citep{Edge_A3667}, 
and is also known to show significant diffuse radio emission around its center \citep{Rottgering_A3667}. 

Abell 4038, also known as Klemola 44, is a rich southern cluster with $z=0.028$,
and is classified as type cD in the Rood-Sastry system, where the cD galaxy is centered at  
$[\alpha(2000)=23^h 47^m 45^{s}.1, \delta(2000)=-28^{\circ}08'26"]$ \citep{Slee_A4038_2001}. 
An X-ray image shows an extended morphology, and there is a radio relic near the 
cD galaxy \citep{Slee_A4038_1998}, 
though it is smaller than the point spread function of typical IACTs (${\sim}0.1^{\circ}$).

The search described in the following sections focused on the detection of point sources
 within the cluster fields
as well as looking for gamma-ray signals from several regions by assuming 
gamma-ray emission models:
The giant radio relics around Abell 3667 may indicate the sites of shocks, 
where particle acceleration
occurs effectively, and thus gamma-ray emission could be expected from the relics 
by the scenario of the UHE proton origin \citep{Inoue_cluster}.
Apart from the shock regions, the density of the ICM is highest at the cluster centers,
so the gamma-ray flux via ${\pi}^{0}$-decay would be strongest there.   

In \S\ref{sec_instrument}, we introduce the CANGAROO-III telescope systems 
and observations of the clusters. 
The data-analysis procedures are explained in \S\ref{sec_analysis}, and the main results 
together with a definition of the gamma-ray search regions are described in \S\ref{sec_result}.
Finally, a discussion of the gamma-ray emission from clusters of galaxies 
based on the CANGAROO-III results is presented. 
Throughout this paper, we assume a Hubble constant of
H$_{0}$=70\,km\,s$^{-1}$\,Mpc$^{-1}$.

\section{Instrument and Observations}
\label{sec_instrument}
Two clusters of galaxies, Abell 3667 and Abell 4038, were observed in the TeV energy band 
using the imaging atmospheric Cherenkov technique with the CANGAROO-III telescope system \citep{Enomoto_Vela}, 
located near Woomera, South Australia (136.786 degree E, 31.099 degree S, 160m a.s.l.).
The system consists of four telescopes, which are located at the corners of a diamond 
with 100~m sides \citep{Enomoto_stereo}.
The first telescope, which we call T1, was not used in these observations, since its current performance 
is inferior to that of the other telescopes. 
The specifications of the second, third, and fourth telescopes,
hereafter called T2, T3, and T4, are almost the same:
they have segmented paraboloid reflector 10 m in diameter and 8 m in focal length. 
Each reflector is composed of 114 spherical mirror facets, and the total
mirror area is 54~m$^2$ \citep{Kawachi_mirror}. 
At each focus there is an imaging camera, which is an array of 427 photomultipliers (PMTs).
Each PMT covers a sky field of $0.17^{\circ}$ in diameter, 
and the total field-of-view is about $4^{\circ}$, suitable for applying analysis using an imaging technique \citep{Kabuki_Camera}.
The data acquisition system is triggered  
when at least two telescopes have signals coinciding for more than 10\,nsec within a
650\,nsec time window, 
thus eliminating muon events that mostly trigger a single telescope \citep{Nishijima_ICRC2005}.
Then, the amplitude and the arrival times of signals from PMTs are digitized by ADC/TDC modules, 
and recorded for off-line analysis \citep{Kubo_Elec_2001}.

The observations were carried out for ON-source and OFF-source tracking runs. 
For OFF-source runs,
the target position was shifted in right ascension so that the telescopes 
tracked the same trajectory
across the sky as ON-source runs.
Also, we adopted {\it wobble mode} observations for both ON-source and OFF-source runs, 
in which the pointing direction was shifted in declination by ${\pm}0.5^{\circ}$ from the target direction every 20 minutes.  
One of the advantages of the {\it wobble mode} is to average the response of PMTs, since the target position rotates on the FOV.
All observations were made on moonless nights from July to September, 2006. 
Details of the data sets are summarized in Table~\ref{tab_observation}. \\
In addition to the observations, we also observed dark regions 
(with no bright stars or gamma-ray sources in the field-of-view)
without imposing the telescope coincidence described above, and we extracted local muon-ring events from these data to 
monitor the total performance of the telescopes \citep{Enomoto_Vela}.
This calibration (denoted {\it muon run}) was done every month, and their statistics are also shown in  
Table~\ref{tab_observation}.
\begin{table}
\caption{Summary of the data set used in this analysis.}
\begin{center}
\begin{tabular}{c c c c c}
\hline
\hline
Cluster& Term & ON\tablenotemark{a} & OFF\tablenotemark{a} & $\langle z\rangle$\tablenotemark{b}  \\
       &      & [hour]              & [hour] & [deg]  \\
\hline
Abell 3667 & Jul-Aug (2006) & 29.7 & 23.7 & 28.6 \\
Abell 4038 & Aug-Sep (2006) & 23.6 & 17.7 & 13.1 \\
{\it muon run}   & Jul-Sep (2006) & 22.4 & -- & 10.8 \\
\hline
\hline
\end{tabular}
\end{center}
\tablenotetext{a}{Dead-time corrected observation time used in our analysis.}
\tablenotetext{b}{Mean zenith angle of ON-source runs.}
\label{tab_observation}
\end{table}

\section{Data Analysis}
\label{sec_analysis}
We followed the analysis procedure explained in detail in \citet{Enomoto_Vela}, 
which we briefly describe here.

First, raw data were calibrated, using daily calibration runs using LEDs \citep{Kabuki_Camera}.
We then selected the shower events from the calibrated datasets.
For each triggered event, camera pixels that recorded less than 5 p.e.\ were discarded so as to remove any night-sky-background photons, 
and shower events were extracted from the remaining pixels, imposing the conditions that 
there were at least 5 adjacent hit pixels,
and that the pixels were triggered within ${\pm}$30\,nsec 
from the mean arrival time of all hit pixels.
This procedure cleaned the shower images and separated random noise, such as multiple night-sky-background photons.
The typical shower event rate was ${\sim} 7$\,Hz (average over 5 minutes), 
and we excluded the data from the analysis when the shower event rate was below
5\,Hz
so as to remove any data affected by clouds etc.
The effective total observation times for the selected datasets, taking account of the dead time in data acquisition, 
are summarized in Table~\ref{tab_observation}.
After this image-cleaning procedure, we discarded events with any hits in the outermost layer of 
the imaging cameras since such shower images may be distorted \citep{Enomoto_RXJ0852}.

Next, image moments ({\it width} and {\it length}) of showers were calculated as defined by \citet{Hillas_moment}, and
the arrival direction of the shower was reconstructed event by event, by
minimizing the sum of the squared \textit{widths} of the images weighted 
by their total photo-electron numbers seen from the assumed direction, 
as described in \citet{Kabuki_CenA}.

Finally, gamma-ray/hadron separation was carried out by applying the Fisher discriminant method 
\citep{Fisher,Enomoto_Vela}. In this method, the Fisher discriminant (hereafter \textit{FD})
 is defined as a linear combination of the image moments,
\begin{equation}
\label{eq_fisher}
\textit{FD} = \sum_{i=1}^{6}{\alpha}_{i}{\cdot}P_{i},
\end{equation}
where $\textbf{P} = (W_{2},W_{3},W_{4},L_{2},L_{3},L_{4})$ is a set of energy-corrected 
\textit{width} and \textit{length} of shower images of three telescopes.
The coefficients, ${\alpha}_{i}$ ($i=1{\sim}6$), were determined so that the difference of the \textit{FD} distribution 
of gamma-ray events and that of hadron events would be maximized.
We used Monte-Carlo simulation data as gamma-ray events, and OFF-source run data as background hadron events
for deriving the coefficients.
We then extracted gamma-ray events from ON-source run data by fitting the ON-source \textit{FD} distribution 
with that of the background (OFF-source) distribution plus the gamma-ray distribution.
Our Monte-Carlo simulation code is based on GEANT3, the details are described in \citet{Enomoto_stereo},
where some parameters such as the geometry of the telescopes are replaced
with those of the current CANGAROO-III system.
The degradation of the overall light collection efficiency 
(including reflectivities of the reflectors, quantum efficiencies of photomultipliers, etc.) 
and the spot size of each telescope were estimated 
from a muon ring analysis \citep{Enomoto_Vela} using {\it muon run} data, 
and they were included in our simulation. 
In the gamma-ray simulation, we assumed a power-law spectrum index of ${\gamma} = -2.1$,
which is often assumed for clusters of galaxies (e.g., \citet{Volk_2000}).

\section{Results}
\label{sec_result}

\subsection{Two dimensional morphology and ${\theta}^{2}$ distribution}
\label{subsec_excess}

\begin{figure*}
\epsscale{1.0}
\plottwo{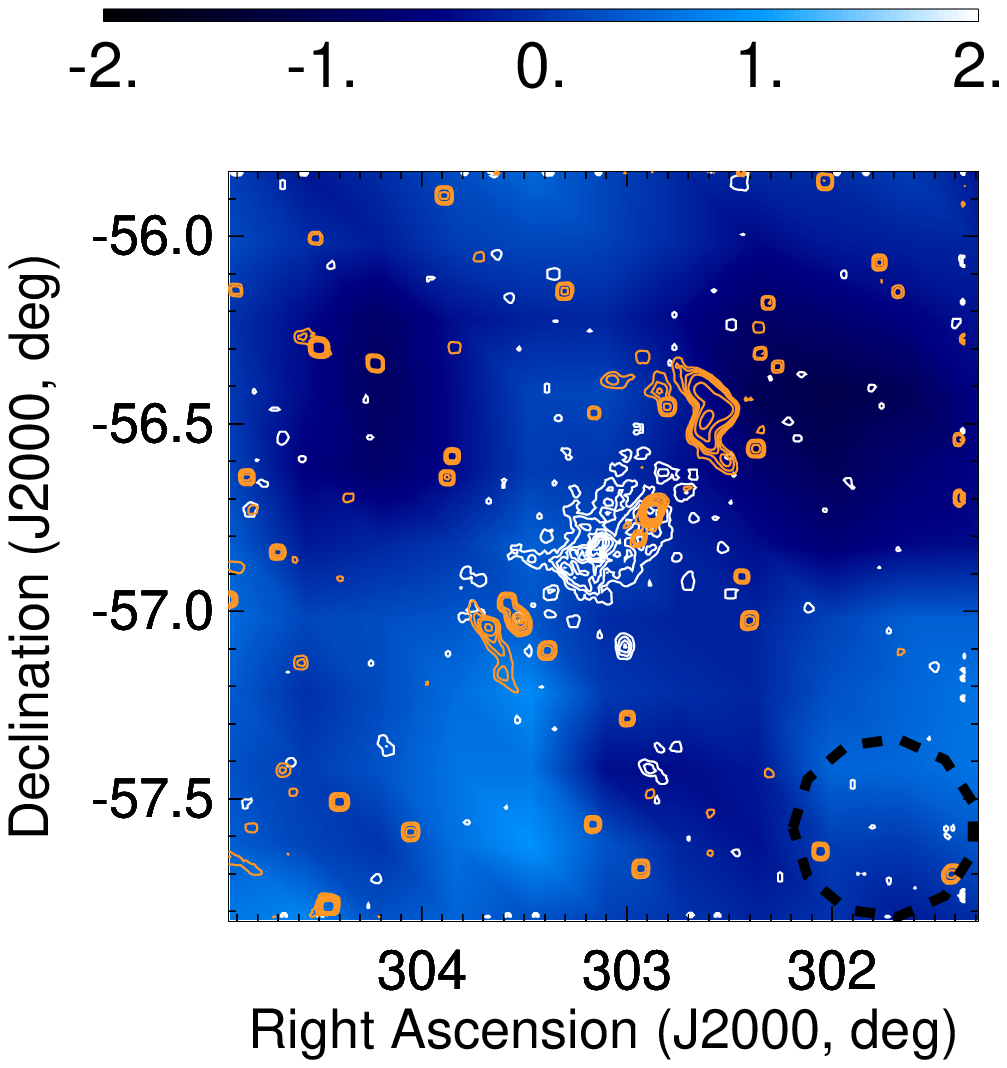}{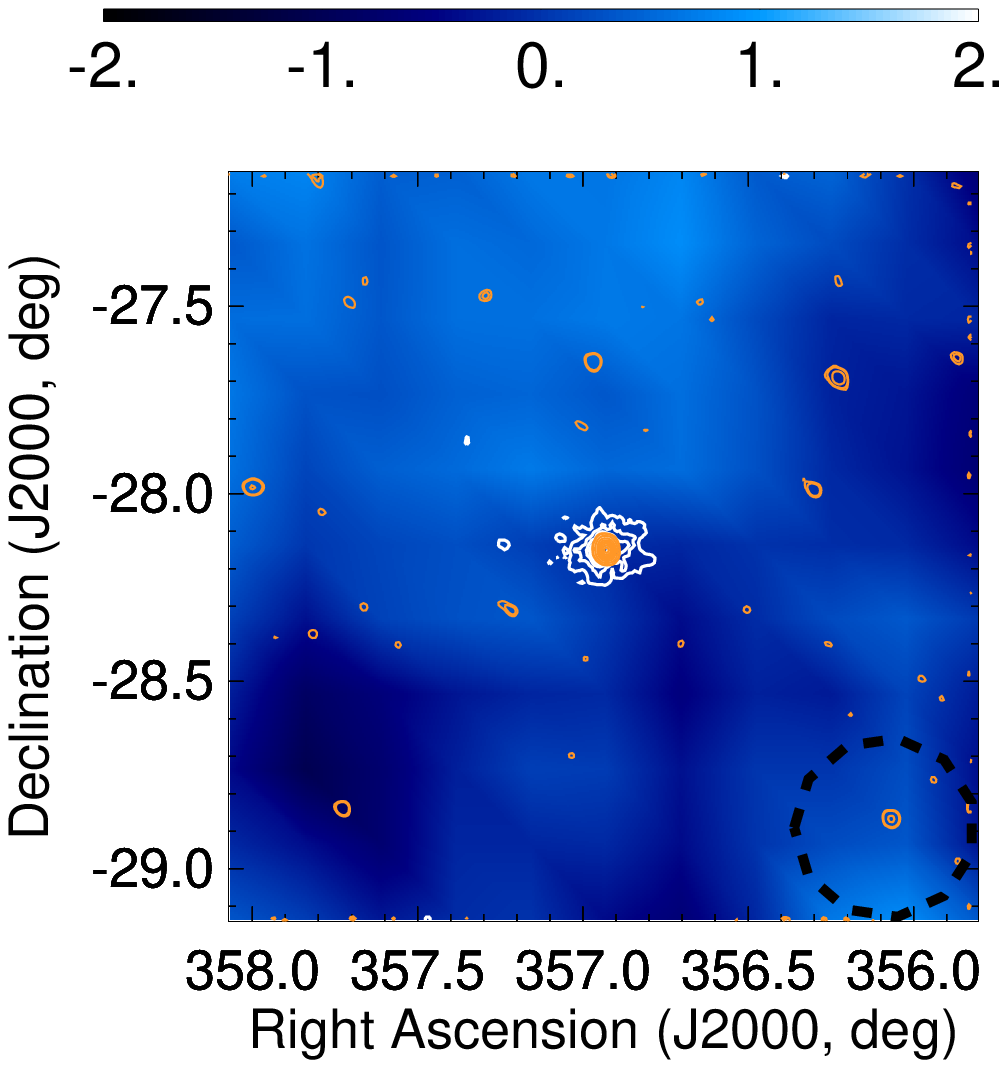}
\caption{Two-dimensional significance maps around the clusters.
The map centers correspond to the position of the cD galaxy of each cluster,
and at each point a smoothing algorithm within $0.6^{\circ}{\times}0.6^{\circ}$ was applied.
Contours of other wavelengths (X-ray \& radio) are over-plotted. 
The left panel is for Abell 3667, where the white contours show \textit{ROSAT} hard-band data \citep{ROSAT_Xray} 
and orange contours show SUMSS 843\,MHz radio data \citep{SUMSS_Radio}.
The right panel is for Abell 4038, where white contours show \textit{ROSAT} hard-band data and 
orange contours show VLA 1.4\,GHz radio data \citep{VLA_Radio}.
Our point-spread function is also shown at the bottom right-hand corner.
}
\label{fig_2Dmap}
\end{figure*}

First, we calculated the two-dimensional (2D) significance map around the cluster centers.
We divided these (ON source) regions into $0.2^{\circ}{\times}0.2^{\circ}$ square bins, 
and calculated the gamma-ray--like excesses and their errors with the \textit{FD} fitting method,
described in the previous section.
Each background (OFF source run) bin was taken so that its position on the field of view would 
correspond to that of the ON region's bin, 
but the area was extended to $3{\times}3$ neighboring bins, to improve the statistical accuracy.  
Fig.~\ref{fig_2Dmap} shows the resulting 2D significance maps of gamma-ray like excesses. 
Since the gamma-ray acceptance falls off toward the outer part of the field-of-view, we limit the map
to within
1~degree from the cluster centers.
The significance distributions from all bins in 2D maps were well approximated by standard normal distributions for both regions.
The best fit Gaussians have mean values of $0.02{\pm}0.10$ (Abell 3667) and $0.17{\pm}0.11$ (Abell 4038), 
with standard deviations of $1.08{\pm}0.08$ (Abell 3667) and $1.02{\pm}0.08$ (Abell
4038), and there are no significant gamma-ray signals.   \\

\begin{figure}
\epsscale{1.0}
\plotone{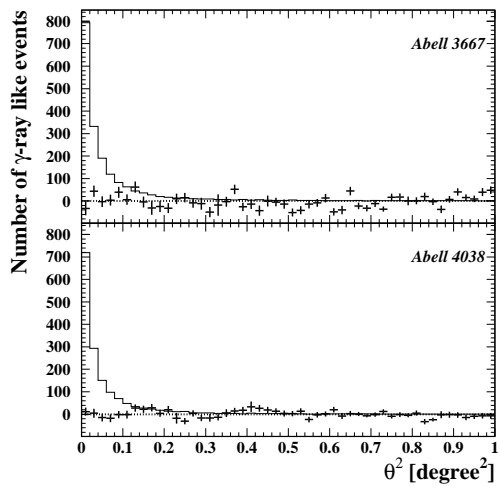}
\caption{${\theta}^{2}$ distributions from cluster centers (where the definition of the center
is the same as in Fig.~\ref{fig_2Dmap}). The upper panel is for Abell 3667 and the lower panel is for Abell 4038. 
The solid histograms show Monte-Carlo simulation results, assuming a
Crab-nebula--level point-like gamma-ray source 
at each cluster center, and
the dotted lines indicate the zero point.
Note that our PSF corresponds to ${\theta}^{2} < 0.06$ degree$^{2}$ . }
\label{fig_theta2}
\end{figure}

Next, we show the ${\theta}^{2}$ distribution, where $\theta$ is the space angle between the target position
and the reconstructed arrival direction, from the cluster centers in Fig.~\ref{fig_theta2}.
The background bin for calculating each ${\theta}^{2}$ bin was taken from the OFF source region
to correspond to the position in the field of view.
Although there were deviations in the ${\theta}^2$ distributions,
they were not significant ($<$3${\sigma}$), considering our point-spread function (${\theta}^{2} < 0.06$ degree$^{2})$.  \\
In summary, there were no detectable point sources in the cluster fields.

\subsection{Gamma-ray emission profiles and upper limits}
\label{subsec_upperlimit}

We also adopted several gamma-ray emission profiles in the cluster fields, and searched for diffuse gamma-ray emission.
First, we defined two circular regions (hereafter \textit{NW}/\textit{SE} \textit{Relic} regions)
that cover the prominent radio relics around Abell 3667, since they may represent
a shock morphology.
The center coordinates (R.A.\ \& Dec.\  in J2000) and their radii were defined as follows: 
$(20^{h}10^{m}24^{s},-56^{\circ}27'00'')$ and $0.30^{\circ}$ for the \textit{NW Relic} region,
$(20^{h}14^{m}36^{s},-57^{\circ}03'00'')$ and $0.24^{\circ}$ for the \textit{SE Relic} region.

We expect that gamma-ray emission via ${\pi}^{0}$-decay is concentrated at
the cluster center regions.
It is well known that many clusters have a diffuse X-ray morphology at their centers,
which may trace thermal components bounded by the gravitational potential of clusters. 
We thus assume that the gamma-ray emission profile traces the X-ray morphology of clusters.
We adopted the \textit{ROSAT} PSPC data for the X-ray morphology.
The peak positions of the X-ray brightness are almost coincident with 
the cD galaxies of the clusters, which were the tracking points of our observations.
We then defined two regions (hereafter \textit{Cluster Core} regions), such that their centers 
were at the position of the cD galaxies  
and the radii were equal to the point where the S/N of the \textit{ROSAT} data fell below ${\sim}$ 3, 
which was $0.40^{\circ}$ for Abell 3667 and $0.26^{\circ}$ for Abell 4038,
as described in Table 2 of \citet{Mohr_xray_radius}.     
For Abell 3667, more recent observations with higher resolution have been reported 
(e.g., XMM-Newton \citep{Briel_xray_A3667};
{\it Chandra} \citep{Vikhlinin_xray_A3667}).
{\it Chandra} has a limited field-of-view for our purpose, however
XMM showed that the signal region above the background noise level was in the central 11$'$ which 
can be regarded as being a point source, considering the positional resolution of CANGAROO-III.
So we first searched for gamma-ray signals from the Abell 3667 center region based on {\it ROSAT} data, 
and we later discuss the case of a point source, especially concerning the cosmic ray energy density.
With the ${\pi}^{0}$-decay model, 
\citet{Volk_2000} assumed that the high-energy protons were accumulated in a cluster through supernova explosions,
and that the predicted proton spectrum forms the power-law index ${\Gamma}=-2.1$ with
an energy cutoff of $E_{max}$=200\,TeV.
So in our Monte-Carlo simulation, gamma-rays were generated uniformly within the defined area
with a power-law spectrum having an index of ${\gamma} = -2.1$.

\begin{figure}
\epsscale{0.85}
\plotone{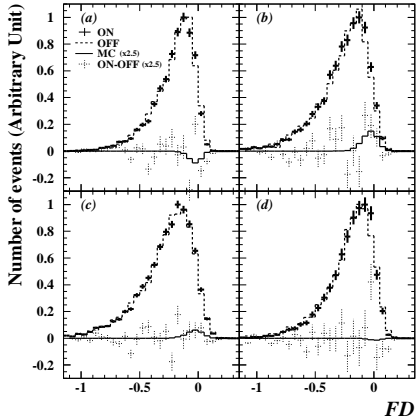}
\caption{{\it FD} distributions from the four regions.
The crosses are \textit{FD} distributions of the ON region, the dashed histograms
are those of the OFF region, the dotted crosses are their subtraction (ON$-$OFF), 
and the solid histogram is that of gamma rays by Monte-Carlo simulation, which were derived from the fitting procedure :
a) the Abell 3667 \textit{NW Relic} region, 
b) the Abell 3667 \textit{SE Relic} region, 
c) the Abell 3667 \textit{Cluster Core} region,
d) the Abell 4038 \textit{Cluster Core} region.
The ON$-$OFF and gamma ray distributions are magnified by 2.5 for clarity.
The resulted gamma-ray flux were at $-$13\%,13\%,13\% and $-$0.9\% of the Crab-level flux for (a)--(d), 
where the negative excesses represent statistical fluctuations, 
and none of the significances exceed 3${\sigma}$. }
\label{fig_fisher_fit}
\end{figure}

The gamma-ray events from each region were calculated by the \textit{FD} fitting method, as before.
The \textit{FD} distribution of each region fitted with that of OFF-source region 
and that of gamma-ray events from a Monte-Carlo simulation is shown in Fig.~\ref{fig_fisher_fit}. 
A gamma-ray signal would appear around \textit{FD}=0
(see, e.g., Fig.1 in \citet{Enomoto_RXJ0852});
however, the calculated significances from the 4 regions did not exceed 3${\sigma}$,
and so there is no evidence of extended emission.

Therefore, we calculated the 2${\sigma}$ upper limits on the integral gamma-ray fluxes from these regions.
The obtained flux upper limits, their threshold energy, and the definition of each region 
are summarized in Table~\ref{tab_upper-limit}.
\begin{table*}
\caption{Summary of results: assumed gamma-ray emission region, threshold energy, and 2${\sigma}$ upper limits.}
\begin{center}
\begin{tabular}{c c c c c c c}
\hline
\hline
Cluster & Region &  \multicolumn{2}{c}{Center}  & Radius & Threshold  & 2${\sigma}$ upper limit  \\
        &        & R.A.(J2000) & DEC(J2000) & [deg] & [TeV] & [cm$^{-2}$sec$^{-1}$] \\
\hline
Abell 3667 & \textit{NW Relic}    & $20^{h}10^{m}24^{s}$ & $-56^{\circ}27'00''$   & $0.30$ & $0.95$ & $3.19{\times}10^{-12}$  \\
           &          &                      &                        &          & $1.45$ & $1.64{\times}10^{-12}$  \\
           &          &                      &                        &          & $2.05$ & $1.05{\times}10^{-12}$  \\
           & \textit{SE Relic}    & $20^{h}14^{m}36^{s}$ & $-57^{\circ}03'00''$   & $0.24$ & $0.85$ & $5.69{\times}10^{-12}$  \\
           &          &                      &                        &          & $1.35$ & $1.86{\times}10^{-12}$  \\
           &          &                      &                        &          & $2.05$ & $1.55{\times}10^{-12}$  \\
           & \textit{Cluster Core} & $20^{h}12^{m}27^{s}.4$ & $-56^{\circ}49'36''$ & $0.40$ & $0.95$ & $5.52{\times}10^{-12}$  \\
           &          &                      &                        &          & $1.35$ & $2.91{\times}10^{-12}$  \\
           &          &                      &                        &          & $1.95$ & $2.12{\times}10^{-12}$  \\
Abell 4038 & \textit{Cluster Core} & $23^{h}47^{m}45^{s}.1$ & $-28^{\circ}08'26''$ & $0.26$ & $0.75$ & $3.30{\times}10^{-12}$  \\
           &          &                      &                        &          & $0.95$ & $2.41{\times}10^{-12}$  \\
           &          &                      &                        &          & $1.35$ & $1.57{\times}10^{-12}$  \\
\hline
\hline
\end{tabular}
\end{center}
\label{tab_upper-limit}
\end{table*}

\section{Discussion}
\label{sec_discussion}

\begin{figure}
\epsscale{1.0}
\plotone{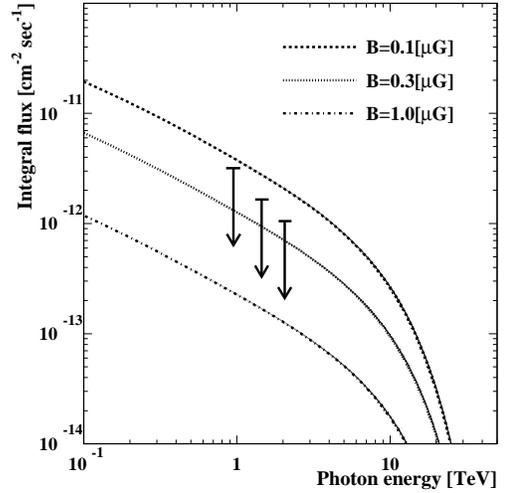}
\caption{Derived gamma-ray flux upper limits from the \textit{NW Relic} region of Abell 3667 
(filled squares) with the predicted gamma-ray fluxes by \citet{Inoue_cluster}.
The model was scaled with the mass and the distance of Abell 3667 to that of a Coma-like cluster, and is shown 
 in the case of magnetic fields of $0.1{\mu}$G, 0.3$\mu$G, and 1.0${\mu}$G. }
\label{fig_integral_flux_Inoue}
\end{figure}

\citet{Inoue_cluster} predicted gamma-ray emission at accretion shocks around a massive cluster. 
We searched for gamma-ray emission from radio relics of Abell 3667, assuming that they might trace
the accretion shock \citep{Ensslin_accretion}, 
although it has also been suggested that the relics are the results of a major merger \citep{Roettiger_A3667_merger}
in which case the particle acceleration would not be as strong as assumed in the
accretion model.
We found no evidence of gamma-ray emission from either region. 
Since estimations of the magnetic field of Abell 3667 have been made for the area of the cluster center and the north-west relic so far
\citep{Johnston_A3667_magnetic},
we compared the derived upper limits from \textit{NW Relic} region with the model prediction, 
as shown in Fig.~\ref{fig_integral_flux_Inoue}.
The model assumes a proton luminosity of one tenth of the kinetic energy flux
through strong accretion shocks, which depends on the cluster mass 
in the form of ${\propto}M^{5/3}$ (see Eq.(2) in \citet{Inoue_cluster}),
and we scaled the predicted gamma-ray flux according to 
the mass (\textit{M}) and distance (\textit{d}) of Abell 3667 
from the (Coma-like cluster) parameters used in their model
(\textit{M}=$2{\times}10^{15}M_{\odot}$, \textit{d}=100\,Mpc).
The mass of Abell 3667 has been estimated using the Virial relation to be
$3.7{\times}10^{15}M_{\odot}$ \citep{Sodre_A3667} or $1.7{\times}10^{15}M_{\odot}$\citep{Girardi_Mass}, 
so here we adopt a cluster mass of their mean value, $2.7{\times}10^{15}M_{\odot}$.
The scaled fluxes are shown in Fig.~\ref{fig_integral_flux_Inoue} with lines 
for magnetic fields of $0.1{\mu}$G, 0.3$\mu$G, and 1.0${\mu}$G. 

Fig.~\ref{fig_integral_flux_Inoue} indicates that we can set a lower limit for 
the magnetic field in the cluster to be $\sim 0.1{\mu}$G, within 
the framework of the model by \citet{Inoue_cluster}.
This value is not a strong constraint on the magnetic field when it is compared with other estimates, 
e.g., a few ${\mu}$G from Faraday rotation measurements (see \citet{Johnston_A3667_magnetic} for other results);
however, the result provides an independent method of a magnetic field estimation, using TeV gamma-ray observations. 
The flux upper limits from the \textit{SE Relic} region were higher than that of the \textit{NW Relic}, 
and the lower limit of the magnetic field strength was 
estimated to be $\sim 0.1{\mu}$G, 
depending on the assumed cluster mass. 
Note that the above flux upper limits also provide a constraint on the gamma-ray emission via primary electron IC emission, 
which is believed to appear at the shocks \citep{Miniati_2003}. 

\begin{figure}
\epsscale{1.0}
\plotone{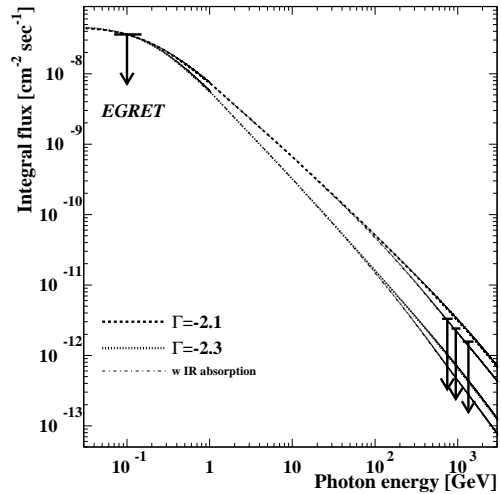}
\caption{Derived gamma-ray flux upper limits from the \textit{Cluster Core} region of Abell 4038
with the gamma-ray emission spectrum via ${\pi}^{0}$-decay process, 
normalized to the EGRET upper limits \citep{Reimer_EGRET}. 
The EGRET upper limits are indicated by arrows at 100~MeV, 
and the lines are the gamma-ray spectrum from the proton power-law indices of $\Gamma=-2.1$ and $-2.3$ 
with an energy cutoff of 200\,TeV.
The gamma-ray absorption effect by IR photons are represented by dot-dash lines,
where the P0.4 model in \citet{Aharonian_EBL} is adopted.  }
\label{fig_integral_flux_Volk}
\end{figure}

We also searched for gamma-ray emission from the \textit{Cluster Core} regions, 
deriving flux upper limits.
The gamma-ray flux via ${\pi}^{0}$-decay, produced in
hadronic collisions of high-energy protons with the ICM,
is thought to be brightest at the cluster centers,
and its flux level is usually discussed from the perspective of 
an effective confinement of cosmic-rays inside a cluster during the Hubble time.
Here, we discuss the total cosmic-ray energy stored inside the cluster centers, using the CANGAROO-III result.
We plotted the flux upper limits from the Abell 4038 \textit{Cluster Core} region 
together with the EGRET upper limit \citep{Reimer_EGRET}
in Fig.~\ref{fig_integral_flux_Volk}.
We adopted the assumptions of \citet{Volk_2000}, as introduced in a previous section,
and the gamma-ray absorption effect by IR photons (P0.4 model in \citet{Aharonian_EBL}) was also incorporated. 
The gamma-ray spectra were represented by lines in Fig.~\ref{fig_integral_flux_Volk},
which were scaled to be consistent with the EGRET upper limits.

As shown in Fig.~\ref{fig_integral_flux_Volk}, the EGRET \& CANGAROO-III results for Abell 4038 
gave almost the same constraint on the gamma-ray emission for the case of ${\Gamma}=-2.1$, and
the total cosmic-ray energy to explain our flux upper limits is
$1.2{\times}10^{63}$\,erg,
using an ICM density of $10^{-3}$ cm$^{-3}$, which is a typical value for cluster centers \citep{Blasi_review}.
We then derived the upper limit of the cosmic-ray energy density within the \textit{Cluster Core} region of Abell 4038,
as ${\sim}40$eV cm$^{-3}$, assuming a spherical symmetry with the radius that we defined for this region.
The same estimation was applied for the \textit{Cluster Core} region of Abell 3667 as well,
and the upper limit was calculated to be ${\sim}20$eV cm$^{-3}$.
Two factors should be considered for this value: 
as described in $\S$\ref{sec_intro}, 
the morphology of Abell 3667 is elongated towards 
two radio relics, rather than a simple sphere, so the assumption of a spherical symmetry might need to be reexamined
where the cluster type of Abell 4038 is cD.
Also, if we adopt the XMM results,
the search region is effectively a point source, as described beforehand.
In this case, a lower flux upper limit 
is obtained, 
and the total volume of the search region also decreased,
with the cosmic-ray energy density increasing to ${\sim}40$eV cm$^{-3}$,
which was the same level for Abell 4038.
In any case, the derived values are 1 order of magnitude higher than that of our Galaxy,
${\sim}1$\,eV\,cm$^{-3}$,
opening the door to discussions of the non-thermal component in the clusters.
All the calculated upper limits are summarized in Table \ref{tab_phys_upper-limit}.

\begin{table*}
\caption{Summary of various limits on the clusters. The defined regions, gamma-ray flux upper limits, lower limit of magnetic field, 
and cosmic-ray energy density}  
\begin{center}
\begin{tabular}{l l c c c c}
\hline
\hline
Region & & 2${\sigma}$ upper limit\tablenotemark{a} & Magnetic field & Energy density \\
       & & [cm$^{-2}$sec$^{-1}$]   & [${\mu}G$]     & [eV cm$^{-3}$]                   \\
\hline
A3667 \textit{NW Relic}    &  & $3.19{\times}10^{-12}$ & $>$0.1 &  -- \\
A3667 \textit{SE Relic}    &  & $5.69{\times}10^{-12}$ & $>$0.1 &  -- \\
A3667 \textit{Cluster Core}& I.\tablenotemark{b} & $5.52{\times}10^{-12}$ & --         &  $<$20 \\
                           & II.\tablenotemark{c} & $2.03{\times}10^{-12}$ & --         &  $<$40 \\
A4038 \textit{Cluster Core}&  & $3.30{\times}10^{-12}$ & --         &  $<$40 \\
\hline
\hline
\end{tabular}
\end{center}
\tablenotetext{a}{The threshold energy was listed in Table \ref{tab_upper-limit}.}
\tablenotetext{b}{Based on \textit{ROSAT} data.}
\tablenotetext{c}{Based on XMM-Newton data (point source analysis).}
\label{tab_phys_upper-limit}
\end{table*}

Further observations of clusters of galaxies 
in the TeV-band by next-generation IACTs currently in the planning stage,
such as CTA\footnote{http://www.mpi-hd.mpg.de/hfm/CTA/} or AGIS\footnote{http://gamma1.astro.ucla.edu/agis/} ,
and in the GeV-band by GLAST\footnote{http://glast.gsfc.nasa.gov/} 
will open a new window in the research of high-energy phenomena of clusters of galaxies
with their improved sensitivities.

\section{Conclusion}
\label{sec_conclusion}

We observed two clusters of galaxies, Abell 3667 and Abell 4038, 
searching for very-high-energy gamma-ray emission with 
the CANGAROO-III atmospheric Cherenkov telescope system in 2006.
No significant excess was detected from the clusters, 
and flux upper limits on the gamma-ray emission were obtained. 
By comparing the upper limit for the north-west radio relic region of Abell 3667
with a model prediction, 
we can derive a lower limit for the magnetic field of the region
of ${\sim}0.1{\mu}$G. 
We also discussed the flux upper limit from the cluster center regions using
a gamma-ray emission model via the decay of ${\pi}^{0}$ produced in hadronic collisions of 
cosmic-ray protons with the ICM.
The upper limit of the cosmic-ray energy density stored within cluster centers 
was estimated to be ${\sim}40$eV cm$^{-3}$ by imposing some assumptions, such as the ICM density,
and the values are 1 order higher than that of our Galaxy.
These estimations show the potential of gamma-ray observations in studies
of the cluster environment.

\acknowledgments

We thank Dr.\ S.\ Inoue for discussions and suggestions on gamma-ray emission from a cluster of galaxies.
This work was supported by a Grant-in-Aid for Scientific Research by the Japan Ministry of Education,
Culture, Sports, Science and Technology, the Australian Research Council, JSPS Research Fellowships,
and the Inter-University Research Program of the Institute for Cosmic Ray Research. We thank the Defense
Support Center Woomera and BAE Systems.

\clearpage

\end{document}